\documentclass[twocolumn]{aastex631}

\shorttitle{Eccentric dust ring formation}
\shortauthors{R. G. Martin \& S. H. Lubow}

\begin{document}

\title{Eccentric dust ring formation in Kozai-Lidov gas disks}

\author{Rebecca G. Martin}
\affil{Department of Physics and Astronomy, University of Nevada, Las Vegas,
4505 South Maryland Parkway, Las Vegas, NV 89154, USA}
\author{Stephen H. Lubow}
\affil{Space Telescope Science Institute, 3700 San Martin Drive, Baltimore, MD 21218, USA}

\begin{abstract} 
A highly misaligned gas disk around one component of a binary star system can undergo global Kozai--Lidov (KL) oscillations for which the disk inclination and eccentricity are exchanged. With hydrodynamical simulations of a gas and dust disk we explore the effects of these oscillations on the dust density distribution. For dust that is marginally coupled to the gas (${\rm St} \approx 1$), we find that  the dust undergoes similar dynamical behaviour to the gas disk but the radial distribution of dust may be very different to the gas. The inward radial drift of the dust is faster in an eccentric disk leading to a smaller outer dust disk radius.  The dust breaks into multiple narrow eccentric rings during the highly eccentric disk phase.  Eccentric dust ring formation  may have significant implications for the formation of planets in misaligned disks.   We suggest that multiple dust rings may generally occur within gas disks that have sufficiently strong eccentricity peaks at intermediate radii. 
\end{abstract} 
  

\section{Introduction}   

Rings of dust in protoplanetary disks are commonly observed \citep[e.g.][]{Vandermarel2019}. The mechanisms behind their formation are still widely debated. The interaction of a planet with a disk is a likely explanation for many rings, although observed rings in systems with ages of $<1\,\rm Myr$ imply that planet formation timescales would need to be short \citep{Helled2014}. Observations of dust rings around Fomalhaut and HD 202628 show that they are are much narrower than the expected width of rings driven by a planet \citep{Faramaz2019,Kennedy2020}.  Alternative explanations include enhanced dust growth at the snow line radius \citep{Zhang2015}.

Some debris disks are observed to have narrow and eccentric rings \citep[e.g.][]{Jayawardhana1998,Koerner1998}. The orbits of the dust grains in an eccentric dust ring are apsidally aligned \citep[e.g.][]{Lee2016}. There is a trend emerging that narrower rings have higher eccentricity \citep{Kennedy2020}.  Eccentric dust rings may be formed by planetary perturbations in the gas disk \citep{lin2019} or without an embedded planet in disks with a high dust to gas ratio \citep{Lyra2013} or by an eccentric cooling instability in self-gravitating disks \citep{Li2021}. 

Disks around young stars that are part of a binary system are frequently observed to be misaligned to the binary orbit \citep[e.g.][]{Jensen2014,Brinch2016}. A highly misaligned disk can be subject to Kozai-Lidov \citep[KL,][]{VonZeipel1910, Kozai1962,Lidov1962} oscillations where the disk inclination and eccentricity are exchanged on a global scale \citep{Martinetal2014b,Fu2015,Fu2015b,Lubow2017,Zanazzi2017}.  The disk communicates radially and undergoes a global response \citep[e.g.][]{Larwoodetal1996}. This instability may have a significant effect on the planet formation process. For example, the eccentricity growth may lead to disk fragmentation \citep{Fu2017}.

Solid material and gas in an accretion disk feel drag forces as a result of their relative velocities \citep{Whipple1972,Adachi1976,Weidenschilling1977b}. Gas in an accretion disk typically orbits at sub-Keplerian velocity because it is partially pressure supported. In the absence of a gas disk, solid particles, on the other hand, orbit at Keplerian velocity. They therefore experience a headwind in the gas disk that causes  them to radially drift inwards. For particles with a size less than about $1\,\rm m$, the gas mean free path is larger than the dust size and the drag force is in the \cite{Epstein1924} regime \citep[e.g.][]{Laibe2012b}. In this regime the fluid is modelled as a collisionless collection of molecules with a Maxwellian velocity distribution and the drag force is proportional to the velocity difference \citep{Armitage2018}.  However, dust and gas that are apsidally misaligned may have a supersonic velocity difference that leads to a gas drag that is quadratic in the velocity difference \citep{kwok1975}. A similar effect may take place in misaligned disks that undergo nodal precession.

The Stokes number is
\begin{equation}
    {\rm St}=\frac{\pi}{2}\frac{\rho_{\rm d}s}{\Sigma_{\rm g}},
    \label{stokes}
\end{equation}
where $\rho_{\rm d}$ is the dust intrinsic density, $s$ is the dust size and $\Sigma_{\rm g}$ is the gas surface density. This is a measure of the coupling between the dust and the gas as a result of the drag force. The strongest radial drift occurs for particles with a size such that ${\rm St}\approx 1$. Dust particles can get trapped in a local pressure maximum in the gas disk \citep{Nakagawa1986}. The radial drift velocity is zero where the pressure gradient is zero \citep[e.g.][]{Rice2006,Zhu2012}.

Previous work on KL disks has considered the evolution of a gas only disk. In this work, for the first time, we consider the effect of the Kozai--Lidov  oscillations on dust in the disk.  Dust that is well coupled to the gas (low Stokes number, St$\,\ll 1$) is forced to undergo  KL oscillations with the gas disk. The dust undergoes KL oscillations on the global disk KL oscillation  timescale. At the other extreme, dust that is completely uncoupled from the gas (high Stokes number, St$\,\gg 1$) undergoes KL oscillations like a test particle. The oscillation time-scale then depends on its distance from the perturbing star. In this case, a misaligned disk of dust  would form a thick annulus as each particle precesses on a timescale that depends upon its distance from the primary star. The particles would only interact with the gas when they passed through the gas disk.     The dust  could also undergo
collisions that act to weaken the KL oscillations and fragment the dust  \citep{Nesvold2016}. We ignore the effects of dust collisions in this work. 
 
In this Letter, we study the interaction of dust with a gas disk that is undergoing KL disk oscillations and becoming highly eccentric. We consider the intermediate regime where St$\,\approx 1$ where there is a relatively strong coupling between the dust and the gas. In Section~\ref{main} we describe the results of  hydrodynamic simulations in which we model a highly misaligned gas and dust disk. We find that the radial dust distribution is strongly affected by the KL oscillations as the dust drifts inwards faster and forms multiple narrow and eccentric dust rings.  We conclude in Section~\ref{conc}.

\section{Hydrodynamical gas and dust disk simulation}
\label{main}

\begin{figure}
\begin{center}
\includegraphics[width=0.43\textwidth]{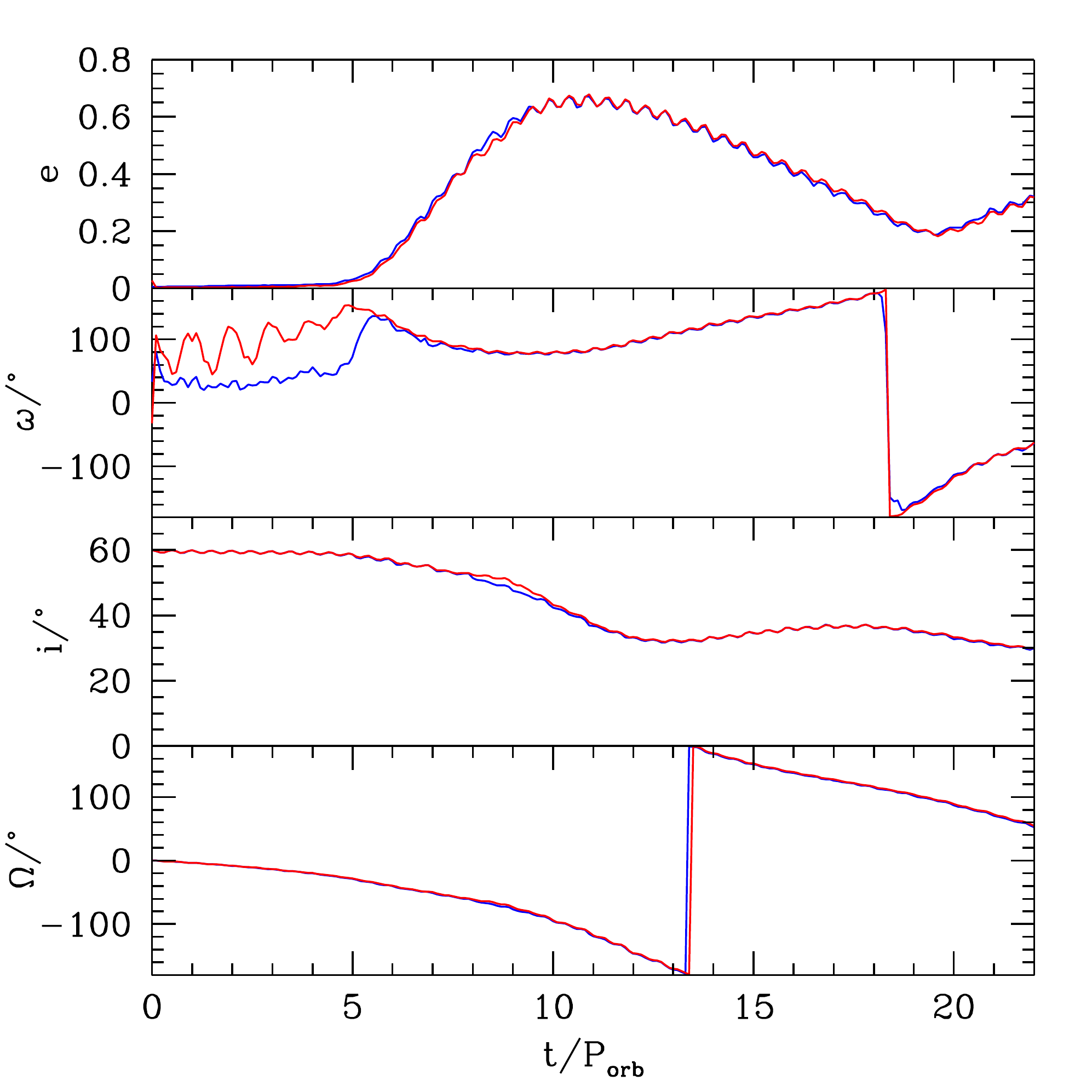}   
\end{center}
\caption{Eccentricity, argument of periapsis, inclination and nodal phase angle evolution of the gas (blue lines) and dust (red lines) in time at a semi-major axis  of $a=20\,\rm au$.   }
\label{eccinc}
\end{figure}

We model a combined gas and dust disk around one component of a binary. In this Section we first describe the simulation setup. Then we discuss the results of a highly misaligned disk that is undergoing KL oscillations. Finally we compare this to simulations of low inclination disks that do not undergo KL oscillations.

\begin{figure}
\begin{center}
\includegraphics[width=0.45\textwidth]{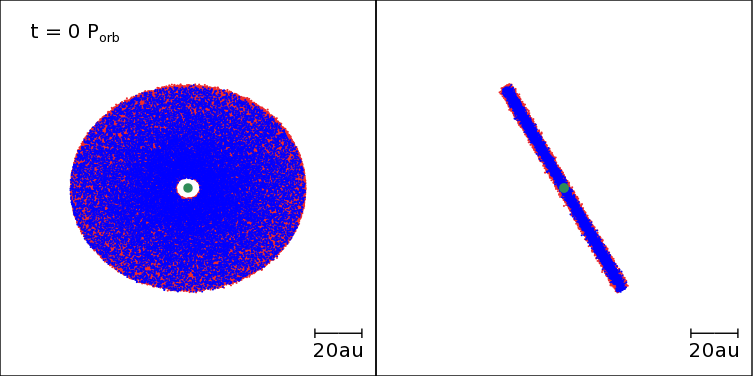}  
\includegraphics[width=0.45\textwidth]{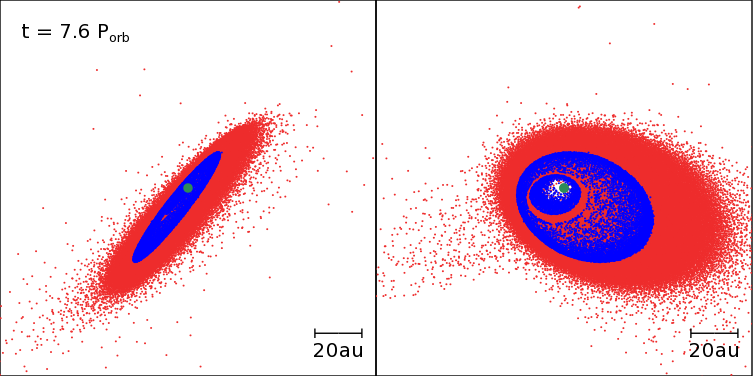} 
\includegraphics[width=0.45\textwidth]{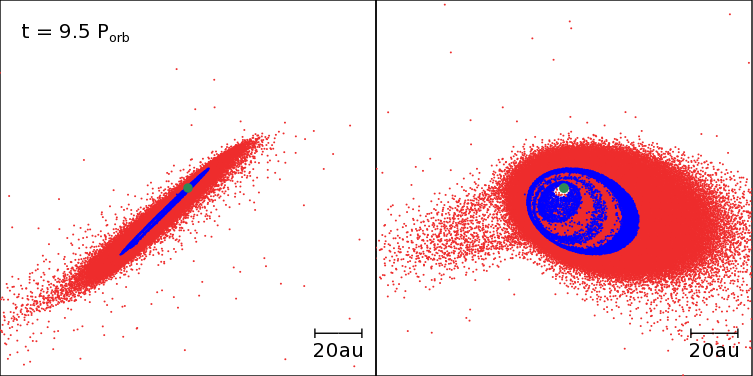}  
\includegraphics[width=0.45\textwidth]{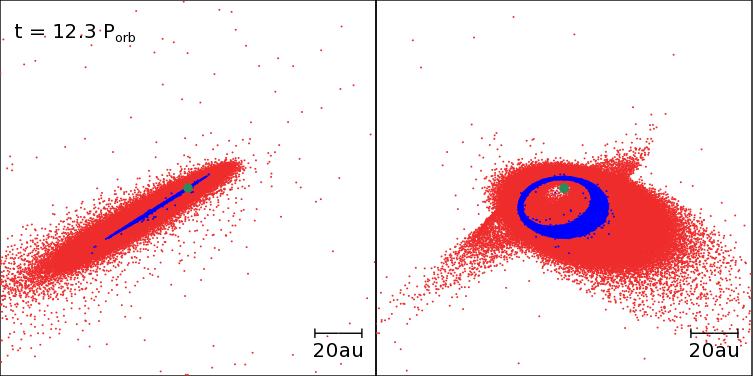}   
\includegraphics[width=0.45\textwidth]{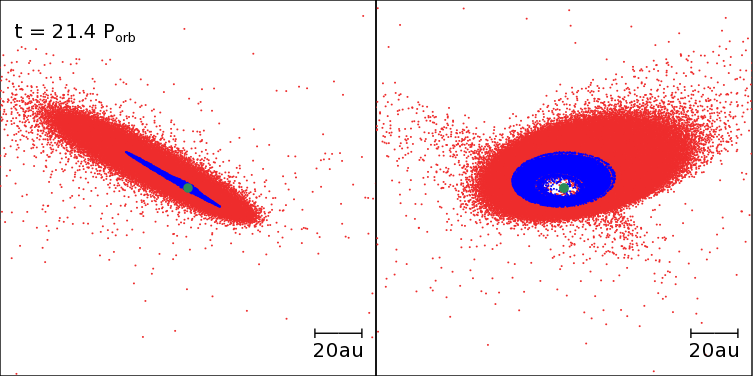} 
\end{center}
\caption{Evolution of the dust and gas disk in time during one KL oscillation. The green circle shows the primary star. The secondary star is along the $x$ axis but is not shown on this scale.  The red points show SPH gas particles while the blue points show the dust particles.  Each row shows the disk from two viewing angles, the $x-z$ plane (left) and the $y-z$ plane (right). The rows show times $t=0$, $7.6$, $9.5$, $12.3$ and $21.4\, P_{\rm orb}$ from top to bottom.   }
\label{fig:main} 
\end{figure}

\begin{figure*}
\begin{center}
\includegraphics[width=0.44\textwidth]{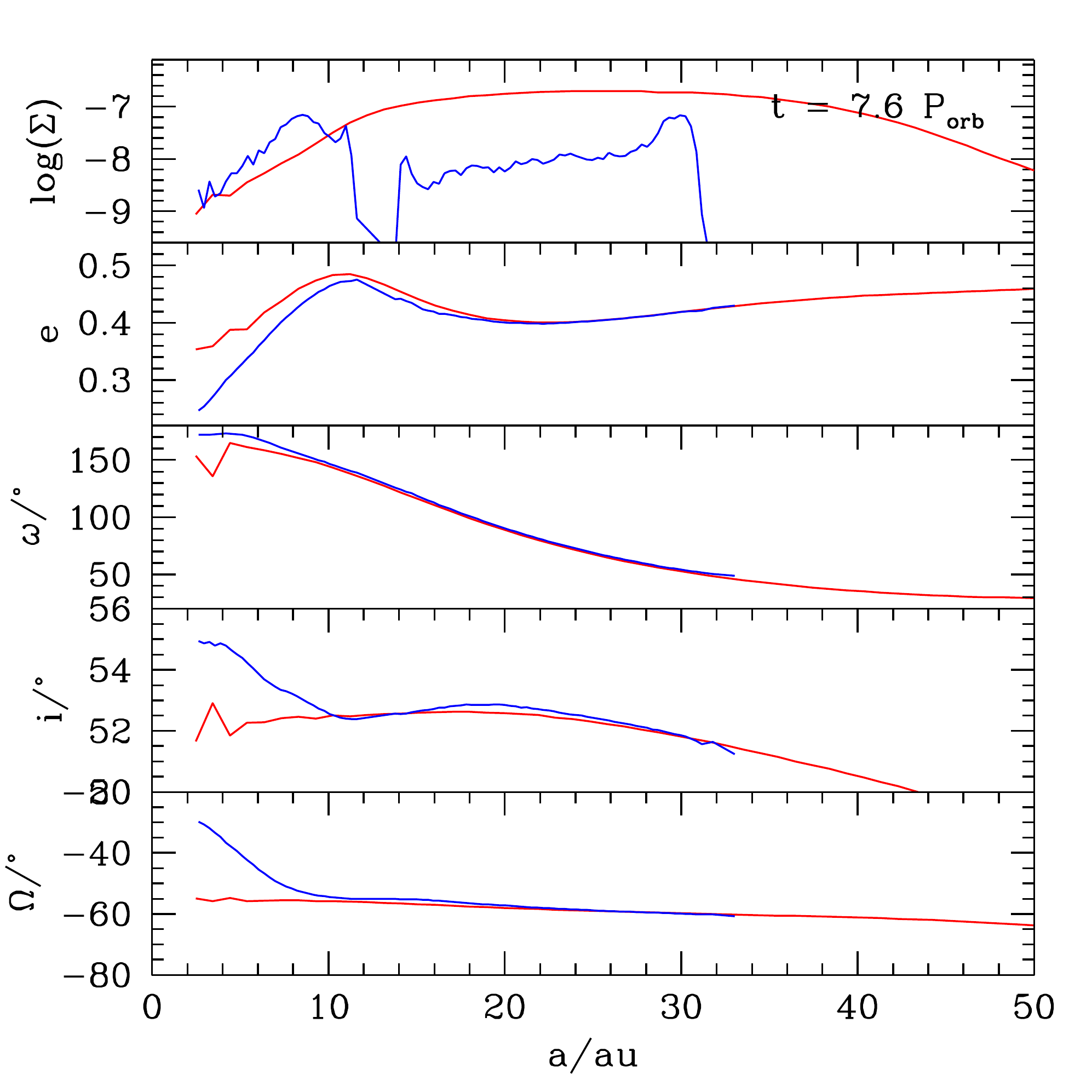} 
\includegraphics[width=0.44\textwidth]{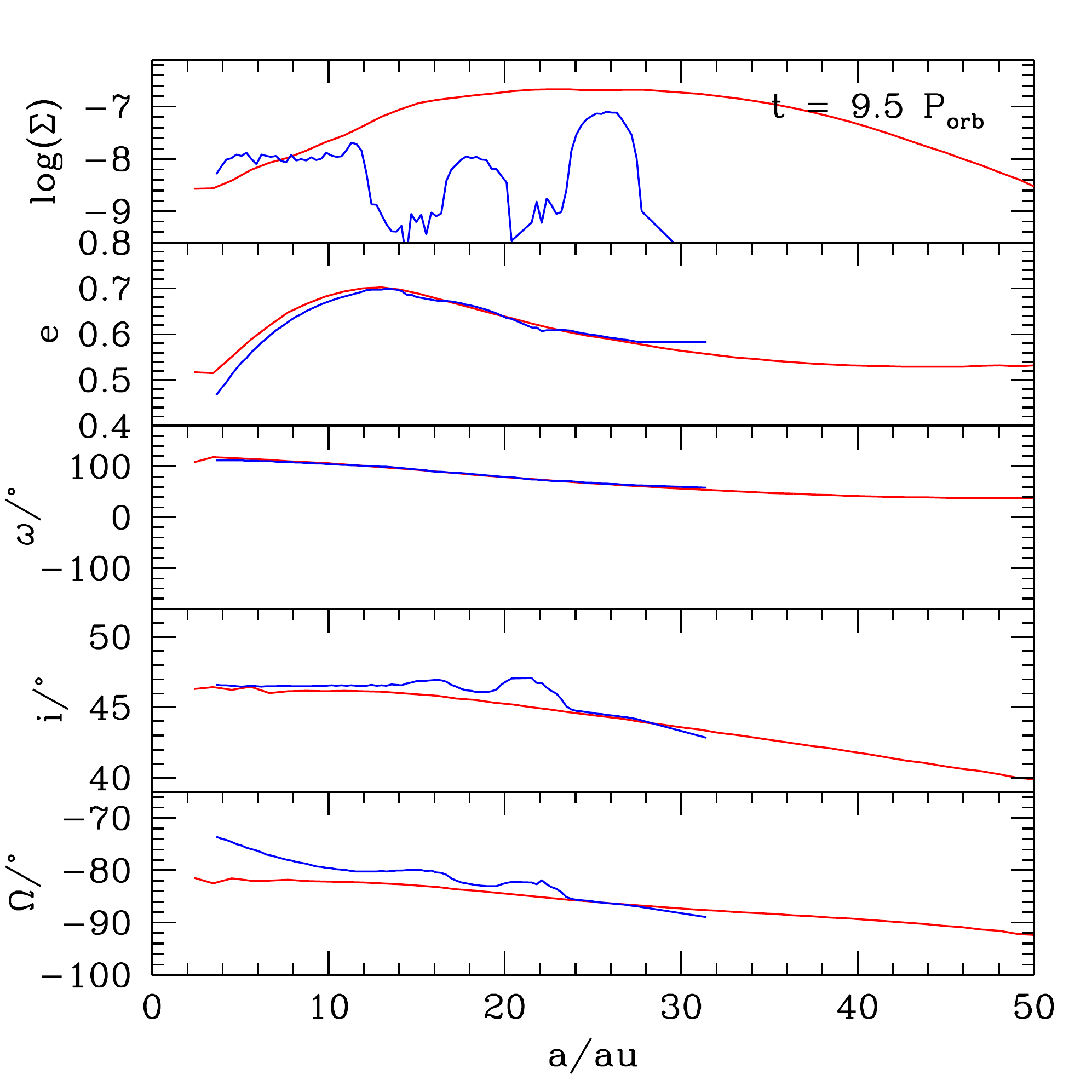}  
\includegraphics[width=0.44\textwidth]{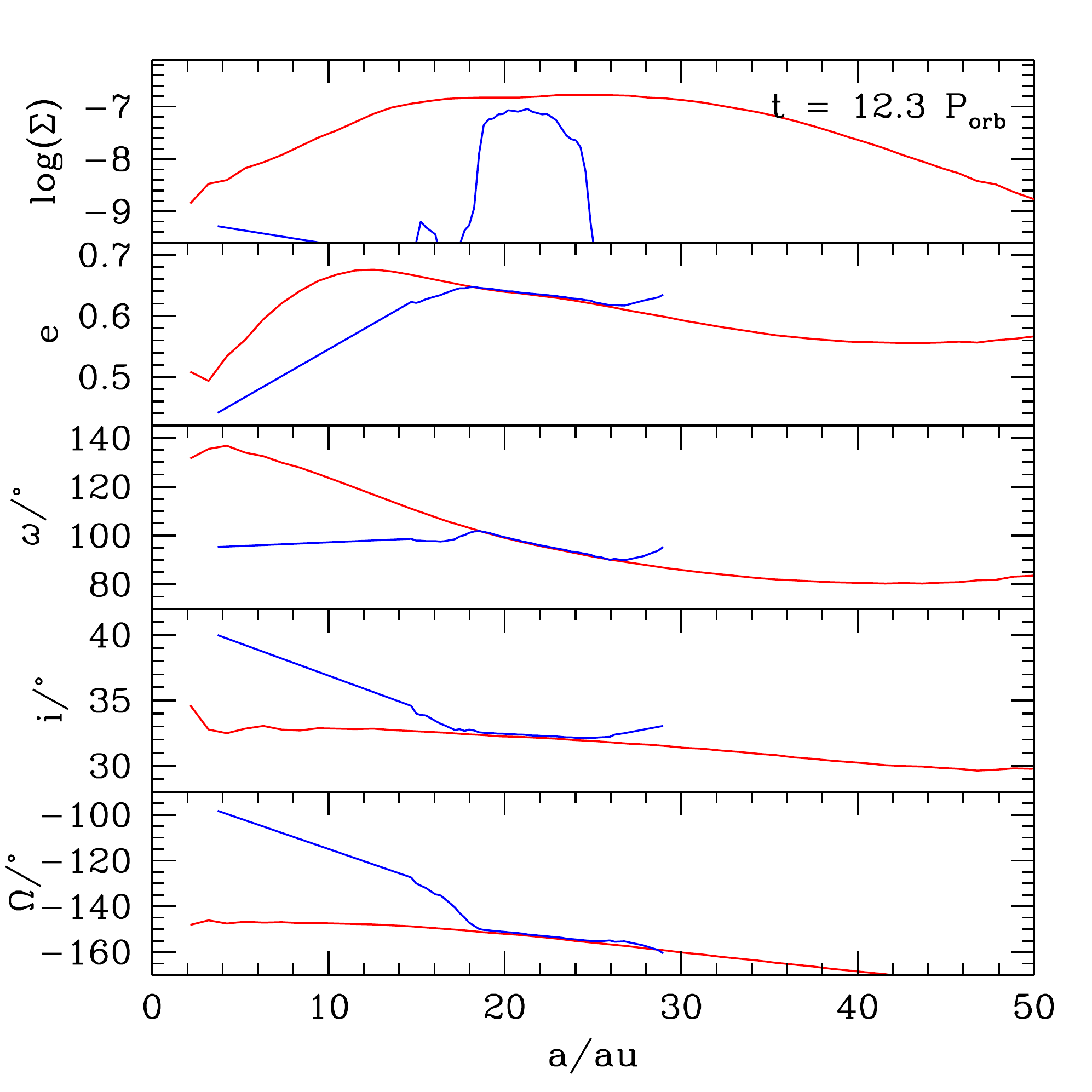}  
\includegraphics[width=0.44\textwidth]{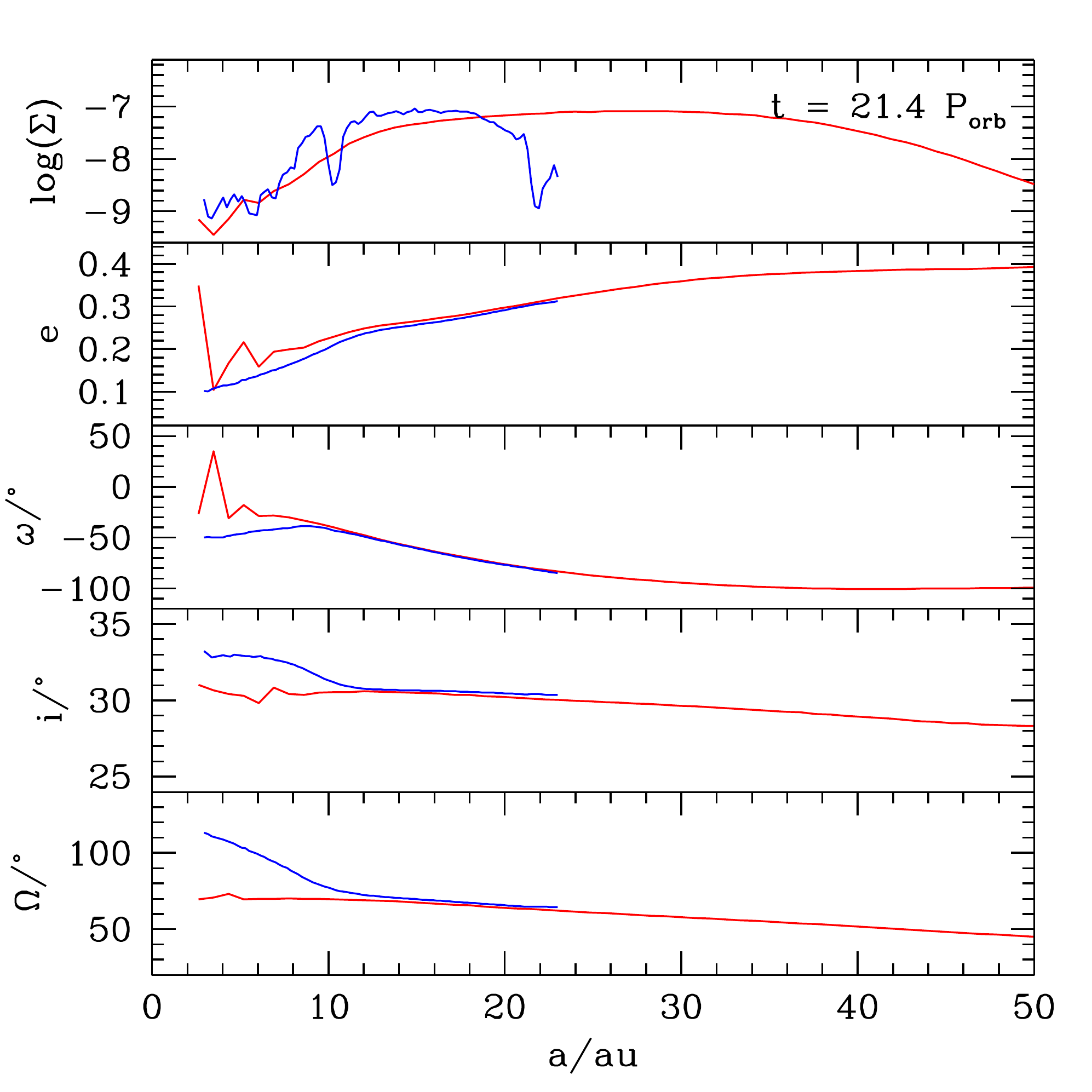}  
\end{center}
\caption{The surface density (in units of $\rm M_\odot \,au^{-2}$), eccentricity, argument of periapsis, inclination and nodal phase angle from top to bottom in each panel at times $t=7.6$ (top left), $9.5$ (top right), $12.3$ (bottom left) and $21.4\, P_{\rm orb}$ (bottom right) for the gas (red lines) and dust (blue lines).   }
\label{fig:sigma} 
\end{figure*} 

\subsection{Simulations setup}

We use the smoothed particle hydrodynamics (SPH) code {\sc Phantom} \citep{PF2010,Price2018} to model a dust and gas disk. We use the two-fluid model in which the dust and gas are treated as different sets of particles that is relevant for large dust particles (with Stokes number St$\,\gtrsim 1$) \citep{Laibe2012,Laibe2012b,Price2015}. The SPH density summation uses the set of nearest neighbours of the same particle type and a drag term in the force equations is added to represent the interaction between the gas and the dust. This method has been tested in this code in many previous works \citep[e.g.][]{Dipierro2015} including disks in misaligned binaries \citep[e.g.][]{Aly2020,Longarini2021}. 

We consider a similar disk setup to \cite{Martinetal2014b}. The binary stars have masses $M_1=M_2=0.5\,\rm M_\odot$ and they orbit in a circular orbit with semi-major axis $a_{\rm b}=200\,\rm au$ with orbital period $P_{\rm orb}=2832\,\rm yr$.  Each star is treated as a sink particle with an accretion radius of $R_{\rm acc}=2\,\rm au$. Any gas or dust particle that moves inside of this radius is accreted and its mass and angular momentum are added to the sink particle \citep[e.g.][]{Bate1995}.   The gas disk has an initial total mass of $0.001\,\rm M_\odot$ and is modelled with $3\times 10^5$ gas particles distributed over $R_{\rm in}=5\,\rm au$ up to $R_{\rm out}=50\,\rm au$ with surface density profile $\Sigma \propto R^{-3/2}$ with a slightly tapered inner edge. 
The disk is gravitationally stable and so we do not include self-gravity in our simulation.  We take the disk aspect ratio to be $H/R=0.035$ at radius $R=5\,\rm au$. The sound speed is locally isothermal with $c_{\rm s}\propto R^{-3/4}$. This ensures that the disk is uniformly resolved \citep{Lodato2010}.

The dust disk is initially distributed in the same manner as the gas disk but with $3\times 10^4$ dust particles initially. The particles have a size of $s=1\,\rm cm$, density of $\rho_{\rm d}=3\,\rm g\, cm^{-3}$ and the dust to gas ratio is $0.01$.  The Stokes number increases with radius from the star in the approximate range of $1-10$.

In order to analyze the results of the simulations, we bin the particles into 100 bins by semi-major axis of the particle orbits. Within each bin we average the orbital properties over the particles such as their eccentricity, inclination etc.

\subsection{Disk undergoing KL oscillations}

We first consider a disk with an initial inclination of $i_0=60^\circ$ to the binary orbital plane, well above the critical inclination required for KL oscillations \citep{Lubow2017,Zanazzi2017}.  Fig.~\ref{eccinc} shows the time evolution of the eccentricity, argument of periapsis, inclination and nodal phase angle  of the gas (red lines) and the dust (blue lines) particles at an orbital semi-major axis of $20\,\rm au$. 
The disk undergoes KL oscillations of eccentricity and inclination, as has been seen previously \citep{Martinetal2014b,Fu2015, Franchini2019}. The dust particles remain relatively well coupled to the gas particles, as evidenced by the fact that the dust and the gas are always within a few degrees of each other.

Fig.~\ref{fig:main} shows the evolution of the gas and dust distributions in time during the course of one KL disk oscillation. We pick the times so that the left panels show the dust disk close to edge on. Initially the disk is circular and inclined by $60^\circ$ to the binary orbital plane (upper panels).  While the (red) gas disk spreads outwards in time through the effect of the viscosity, the (blue) dust undergoes inward radial drift and clears out from the outer parts of the disk.  Notice that in the right panels there is a faint gas disk in the background that is misaligned with respect to the main disk. This additional disk is a consequence of disk mass transfer to the companion star \citep{Franchini2019}. The disk eccentricity begins to grow and peaks at a time of around $12\,P_{\rm orb}$ (see Fig.~\ref{eccinc}). Fig.~\ref{fig:sigma} shows the radial profile of the disk surface density and orbital elements as a function of semi-major axis of the particles at the same times as in Fig.~\ref{fig:main}. Initially, as the dust moves inwards, there is a local peak in the dust distribution that grows in the inner parts of the disk at a semi-major axis of around $6-8\,\rm au$. This is a result of the pressure bump in the inner parts of the disk caused by the distribution of the gas.  

The second row in  Fig.~\ref{fig:main} shows the disk during the initial time of eccentricity growth.  At at time of $7.6\,P_{\rm orb}$ there are two disjoint dust rings.  The eccentricity of the disk does not grow completely uniformly with radius (as seen in the top left panel of Fig.~\ref{fig:sigma}).  Two rings form when the eccentricity profile has a local peak. This occurs at a semi-major axis of $10\,\rm au$ at a time of $7\, P_{\rm orb}$. The outer edge of the innermost ring is at the local peak in the eccentricity profile. This radius moves slightly outwards in time and the outer edge of the innermost dust ring follows this.  By a time of $9.5\,P_{\rm orb}$, the outer edge of the inner most ring has moved outwards to about $12\,\rm au$, as seen in the top right panel of Fig.~\ref{fig:sigma}. 

Three rings are present at a time of $9.5\,P_{\rm orb}$ (see the third row in Fig.~\ref{fig:main}). The ring structure is a result of velocity differences between the dust and the gas. This could arise from differences in eccentricity, argument of periapsis, inclination, or nodal phase angle.  Ring formation has been seen previously in misaligned circumbinary disk simulations as a result of nodal precession. Where the velocity difference between the dust and the gas is zero, there is no radial drift and so dust piles up \citep{Longarini2021}.

The innermost ring gets accreted on to the star and by a time of about $10.5P_{\rm orb}$ there is only one dust ring in the simulation. The fourth row of Fig.~\ref{fig:main} and the lower left panel of Fig.~\ref{fig:sigma} shows the disk at a time of $12.3\,\rm P_{\rm orb}$ where there is one dust ring. This narrow and eccentric dust ring is long lived. But as the disk becomes more circular again, the dust spreads inwards and reaches the inner edge of the disk at a time of about $18\,P_{\rm orb}$. At this time the disk reaches its lowest eccentricity during the KL oscillation. The lowest panels in Fig.~\ref{fig:main} and the lower right panel of Fig.~\ref{fig:sigma} show the dust disk has extended inwards to the central star at a time of $21.4\,\rm P_{\rm orb}$. 

We have only shown the evolution during one KL disk oscillation, but the behaviour may repeat over time. If there is a source of high inclination material accreting on to the disk, then the KL disk oscillations may be long lived \citep{Smallwood2021}. For the simulation presented in this Section, the first KL oscillation occurs over a timescale of about $20\,\rm P_{\rm orb} \approx 57,000\,\rm yr$. One to three narrow dust rings are present from a time of about $7.6\,P_{\rm orb}$ until $18\,P_{\rm orb}$. Thus, dust rings are present for roughly half the time of the first oscillation. If strong oscillations are long lived, then we may expect dust rings to be present transiently for half the disk lifetime. However, if the KL oscillations damp out over time, then dust rings may only be present for a short fraction of the disk lifetime. Note that a wider orbit binary drives KL oscillations on a longer timescale.

\begin{figure}
\begin{center}
\includegraphics[width=0.5\textwidth]{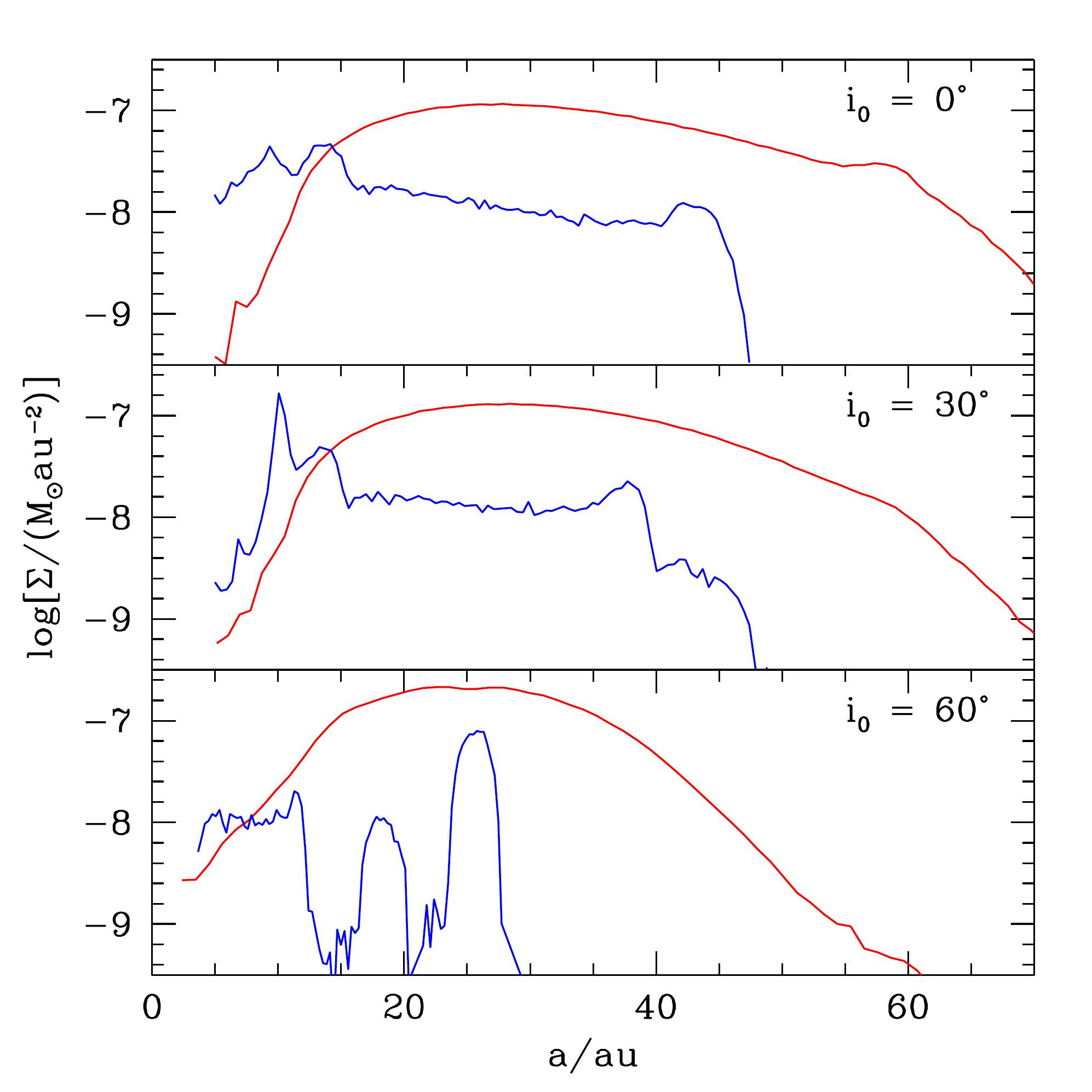}  
\end{center}
\caption{Surface density profile for the gas (red lines) and dust (blue lines) for three different simulations at a time of $9.5\,P_{\rm orb}$. The upper panels the disk is coplanar, in the middle panels $i_0=30^\circ$ and in the lower panels $i_0=60^\circ$ (this is the same as the top right panel in Fig.~\ref{fig:sigma}). Note that the upper two simulations have a sink size of $5\,\rm au$ while the lower one has a sink size of $2\,\rm au$. }
\label{comparison} 
\end{figure} 

\subsection{Outer edge of the dust disk}

In order to disentangle the effects caused by the KL disk oscillations we now consider two additional simulations with different initial disk inclinations: coplanar to the binary orbit and $i_0=30^\circ$. In the interests of computational time, the sink size in these simulations is $5\,\rm au$, otherwise everything else about the disk setup is the same. Thus, differences in the inner parts of the disk may be a result of the sink size. However, differences in the outer parts of the disk are a result of the varying disk inclinations.  Fig.~\ref{comparison} shows the surface density profiles for the disks at a time of $9.5\,P_{\rm orb}$. We also include for comparison the disk undergoing KL oscillations (same as the surface density in the top right panel in Fig.~\ref{fig:sigma}).  In the coplanar and $i_0=30^\circ$ simulations the disk does not become significantly eccentric. A disk that is not eccentric is able to extend to much larger semi-major axis. Furthermore, there are no breaks in the dust distribution. The dust is still cleared from the outer parts of the disk, but not so significantly as in the eccentric disk. The outer radius of the dust disk is similar in the coplanar and $i_0=30^\circ$ disks. Note that radial drift is already thought to be faster in a disk in a binary compared to a single star disk \citep{Zagaria2021} and the disk eccentricity speeds this up even further.

\section{Discussion}

The formation of eccentric dust rings depends upon the Stokes number (equation~\ref{stokes}) that is sensitive to the dust grain size and the gas disk surface density. For small grains, the dust remains tightly coupled to the gas disk and will not display any ring structure. For large solid bodies, they will behave independently of the gas and form a thick annulus. We have also considered some simulations with different dust grain sizes (but otherwise the same disk setup) with a disk inclination of $i_0=60^\circ$. For a simulation with a smaller dust grain size of $1\,\rm mm$,  the ring behaviour is not so pronounced. The inner ring breaks off, but there is not a period of time with more than two dust rings in the disk. This is because the solid particles are more tightly coupled to the gas. For a larger size of $10\,\rm cm$, again we see two dust rings but not three.  However, if the disk was more massive, the Stokes number for a fixed particle size would be smaller. Thus, the more massive the disk, the larger the particles for which we expect the eccentric dust ring formation to occur.  
While the ring structure may form in a disk made up entirely of small particles, small particles may also be formed as a result of collisions between larger particles that have the ring structure. The KL oscillations of the disk lead to shocks within the gas \citep{Fu2017} and also likely to lead to collisions between solid bodies. Collisions are more likely to occur near pericentre thus leading to a brightness asymmetry in some debris disks \citep{Pan2016,Olofsson2019}. 

ALMA is sensitive to observing sub-mm size grains \citep[e.g.][]{Zhu2019b}. This is somewhat smaller than the size of the particles that we have found to form eccentric dust rings. However, eccentric dust rings formed through an eccentric gas disk may be observable if larger particles that are not observable) form a ring structure and collisions between them lead to the formation of smaller dust particles that are observable.
The relative width of the outermost ring  at time of $9.5\,P_{\rm orb}$ is $\Delta r/a \approx 4/26=0.15$. This is relatively narrow, but, similar to the widths measured in the rings in  Formalhaut and HD 202628. At a time of $12.3\,P_{\rm orb}$ the width of the single ring is $\Delta r/a \approx 6/22=0.27$. While the ring widths in Formalhaut and HD 202628  are of a similar magnitude to our simulations, the magnitude of the eccentricity of the observed dust rings are around 0.12, much smaller than that seen in our simulations. Thus we do not think that the KL mechanism is driving the eccentricity growth in those systems, but the dust ring formation may be a general outcome involving an eccentric gas disk. The dust ring formation presented in this work may occur for gas disks with an eccentricity driven by mechanisms other than KL oscillations. 

Because of the strong radial drift, there are no dust particles with St close to 1  near the outer edge of the disk in a binary. Material that is transferred to the disk around the companion star or to a circumbinary disk during the KL oscillations \citep{Franchini2019} contains only gas and very small grains with ${\rm St} \ll 1$. No large dust particles are transferred. This can be seen in the right panels of Fig.~\ref{fig:main}. The small mass disk around the secondary star can be seen behind the primary disk. This has implications for planet formation in that the material that is transferred to the companion or the circumbinary disk may increase the ratio of gas to dust in that disk. The fast radial drift also means that the size of highly misaligned dust disks around one component of a binary may be even smaller compared to the size of a coplanar disk.

A misaligned disk of planetesimals around one component of a binary, in the absence of gas, undergoes nodal precession that leads to the particles being distributed in a thick annulus. The relative velocities between particles may lead to fragmentation rather than accretion and planet formation is difficult, especially at inclinations where the particles undergo KL oscillations \citep{Marzarietal2009}. Similarly, relative velocities may be increased where eccentricity is driven in the particles and they apsidally precess \citep{Thebault2008}. However, we have shown that some particles may be able to form dust rings that are in apsidally aligned orbits and this alignment reduces the relative velocities \citep{Marzari2000}. When dust rings form, the dust becomes compressed into a radially narrow region and so in the ring, the dust to gas ratio may approach 1 (see Fig.~\ref{fig:sigma}). This could lead to efficient planetesimal formation via the streaming instability \citep{Youdin2005,Yang2017}. Recent calculations have shown that the streaming instability timescale may be increased by the presence of multiple dust species \citep{Krapp2019,Zhu2021,Yang2021}. A misaligned/eccentric gas disk may provide a favourable environment since only a narrow range of dust sizes may remain in the disk.  Furthermore, planet formation may be able to proceed at inclinations above the critical KL angle.

\section{Conclusions}
\label{conc}

A highly misaligned disk around one component of a binary star system undergoes global KL oscillations of the disk eccentricity and inclination.  Dust that is sufficiently well coupled to the gas undergoes KL oscillations on the same timescale as the gas disk. However, the radial dust distribution may be very different to the gas. The dust has a faster inward radial velocity as a result of the disk eccentricity and the dust may break into disjoint rings during periods of high disk eccentricity.

\section*{Acknowledgements} 
We thank an anonymous referee for useful comments.  We thank Daniel Price for providing the phantom code for SPH simulations and acknowledge the use of splash \citep{Price2007} for the rendering of the figures. Computer support was provided by UNLV's National Supercomputing Center. We acknowledge support from NASA through grants 80NSSC21K0395 and 80NSSC19K0443.

\bibliographystyle{aasjournal}
\bibliography{ms}

\end{document}